\documentstyle[prl,aps]{revtex}
\def\rvec{\vec{r}}

\def\half{{1\over 2}}
\def\Ekin{E_{\rm kin}}
\def\Eho{E_{\rm ho}}
\begin{document}
\title{Sum Rule Approach to Collective Oscillations
of Boson-Fermion Mixed Condensate of Alkali Atoms}
\author{T. Miyakawa, T. Suzuki and H. Yabu}
\address{Department of Physics, Tokyo Metropolitan University, 
         1-1 Minami-Ohsawa, Hachioji, Tokyo 192-0397, Japan}
\date{\today}
\maketitle
\begin{abstract}
The behavior of collective oscillations of a trapped boson-fermion mixed 
condensate is studied in the sum rule approach. Mixing angle of 
bosonic and fermionic multipole operators is introduced so 
that the mixing characters of the low-lying collective modes are 
studied as functions of the boson-fermion interaction strength. 
For an attractive boson-fermion interaction, the low-lying monopole 
mode becomes a coherent oscillation of bosons and fermions and shows 
a rapid decrease in the excitation energy towards the instability point of 
the ground state. In contrast, the 
low-lying quadrupole mode keeps a bosonic character 
over a wide range of the interaction strengths.  For the dipole mode 
the boson-fermion in-phase oscillation remains to be the eigenmode 
under the external oscillator potential. For weak repulsive values of 
the  boson-fermion interaction strengths we found that 
an average energy of the out-of-phase dipole mode stays lower 
than the in-phase oscillation. 
 Physical origin of the behavior of the multipole modes against 
boson-fermion interaction strength is discussed in some detail. 
\end{abstract}
%

Collective oscillation is one of the most prominent phenomena 
common to a variety of many-body systems. The realization of 
the Bose-Einstein condensates (BEC) for trapped Alkali 
atoms\cite{review} offers a possibility to study such phenomena 
of quantum systems under ideal conditions. Up to now the 
experimental\cite{Coll} as well as theoretical\cite{Stringari,Kimura,TBEC} 
studies of collective motions of BEC have been intensively performed. 
Quite recently the degenerate Fermi gas of trapped $^{40}$K atoms 
has been realized\cite{DeMarco}, which motivates the study of collective 
motion also in Fermi gases\cite{TFex}. These developments further 
encourage the study of  possible boson-fermion mixed condensates 
of trapped atoms\cite{symbf,hydro}. Now the static 
properties\cite{Molmer,Nygaard,Miyakawa}, stability 
conditions\cite{Minniti,Instbf}, and some 
dynamical properties\cite{Amoruso,tsurumi} of trapped boson-fermion 
condensates have been investigated. In the present paper, we study 
the behavior of collective oscillations 
of a boson-fermion mixed condensate at $T=0$ for both repulsive and 
attractive boson-fermion interactions. We adopt the sum rule 
approach\cite{Stringari,Kimura}  that has proved to be successful 
in the studies of collective excitations of Bose condensates. 
For the mixed condensate, we  introduce a mixing angle 
of the boson/fermion excitation operators  so as to allow the 
in- and out-of-phase oscillations of the bose and fermi condensates.


In the sum rule approach we first calculate the energy weighted 
moments $m_p=\sum_{j}(E_{j}-E_{0})^{p}|\langle j|F| 0 \rangle|^{2}$ 
of the relevant multipole operator $F$, where  $|j\rangle$'s represent 
the complete set of eigenstates of the Hamiltonian with energies $E_{j}$, 
and $|0\rangle$ denotes the ground state. The excitation energy 
is expressed as $\hbar\omega=(m_3/m_1)^{1/2}$ which provides a useful 
expression for the average  energy of the collective 
oscillation\cite{Stringari,TFex}. 
The moments are calculated from formulae 
$m_{1}=\half \langle 0| \left[F^{\dagger},[H,F]\right] |0\rangle$ and 
$m_{3}=\half \langle 0| \left[[F^{\dagger},H],\left[H,[H,F]\right]
\right] |0\rangle$. We consider three types of multipole operators which 
are defined by 
\begin{equation}
   F_\alpha^\pm \equiv \sum_{i=1}^{N_b}f_\alpha(\rvec_{bi})\pm 
   \sum_{i=1}^{N_f} f_\alpha(\rvec_{fi}),\quad (\alpha=M,D,Q)
\end{equation}
where the functions $f_\alpha$ are defined by  $f_M(\rvec)=r^2$ for 
monopole, $f_D(\rvec)=z$ for dipole,  and  $f_Q(\rvec)=3z^2-r^2$ for 
quadrupole excitations. The 
indices $b,f$ denote boson/fermion, $N_b, N_f$ the numbers of 
bose/fermi particles, and $\pm$ correspond to 
the in-phase and out-of-phase oscillation of the two types of particles. 
We actually take a linear combination of the form 
\begin{equation}
    F_{\alpha}({\bf r};\theta) =F^+_{\alpha} \cos{\theta}
    +F^-_{\alpha}\sin{\theta} \qquad (-\frac{\pi}{2} < \theta\le \frac{\pi}{2})
\end{equation}
parametrized by the mixing angle $\theta$.  We study the value of $\theta$ 
that minimizes the calculated frequency $\omega$ for each $\alpha$. 

We consider the polarized boson-fermion mixed condensate 
in spherically symmetric harmonic oscillator potential. 
The system is described by the Hamiltonian
\begin{equation}
\label{bfH}
    H = \sum_{i=1}^{N_{b}} \left\{\frac{{\vec{p}}\,^{2}_{bi}}{2m}
    + \half m\omega_0^{2}\rvec\,^{2}_{bi}
    + \half g \sum_{j=1}^{N_{b}}\delta(\rvec_{bi}-\rvec_{bj}) \right\}
    + \sum_{i=1}^{N_{f}} \left\{ \frac{{\vec{p}}\,^{2}_{fi}}{2m}
    +\half m\omega_0^{2}\rvec\,^{2}_{fi}\right\}
    + h \sum_{i=1}^{N_{b}}\sum_{j=1}^{N_{f}}\delta(\rvec_{bi}-\rvec_{fj})
    \end{equation}
where we assume the same oscillator frequencies and masses for bosons and 
fermions for simplicity. 
The coupling constants $g,h$ are the boson-boson/boson-fermion 
interaction strengths represented by the $s$-wave scattering lengths 
$a_{bb}$ and $a_{bf}$ as $g=4\pi\hbar^{2}a_{bb}/m$
and $h=4\pi\hbar^{2}a_{bf}/m$. The fermion-fermion interaction has been neglected 
as the polarized system is considered. Following the standard calculation 
procedure the excitation frequencies are obtained as:

(i) Monopole
\begin{equation}
\label{egm}
\frac{\omega_{M}(\theta)}{\omega_0}=\sqrt{2}\sqrt{1+\frac{\Ekin^{+}
+\frac{9}{4}E_{bb}+\frac{9}{4}E_{bf}+2(\Ekin^{-}+\frac{9}{4}E_{bb}
+\frac{3}{4}\Delta')\cos{\theta}\sin{\theta}
-\Delta\sin^2{\theta}}
{\Eho^{+}+2\Eho^{-}\cos{\theta}\sin{\theta}}}
\end{equation}

(ii) Quadrupole
\begin{equation}
\label{egq}
\frac{\omega_{Q}(\theta)}{\omega_0}=\sqrt{2}\sqrt{1+\frac{\Ekin^{+}
+2 \Ekin^{-}\cos{\theta}\sin{\theta}-\frac{2}{5}\Delta \sin^2{\theta}}
{\Eho^{+}+2\Eho^{-}\cos{\theta}\sin{\theta}}}
\end{equation}

(iii) Dipole
\begin{equation}
\label{egd}
\frac{\omega_{D}(\theta)}{\omega_0}=\sqrt{1-\frac{4}{3\hbar\omega_0}
\frac{\Omega \sin^2{\theta}}
{N^{+}+2 N^{-}\cos{\theta}\sin{\theta}}}
\end{equation}

Here we defined $\Ekin^{\pm}\equiv \Ekin^{b}\pm \Ekin^{f},
\Eho^{\pm}\equiv \Eho^{b}\pm \Eho^{f}$ and $N^{\pm}\equiv 
N_{b}\pm N_{f}$, where $\Ekin^{\{b,f\}}$ and $\Eho^{\{b,f\}}$ 
are respectively the expectation values of the kinetic and oscillator 
potential energies for boson/fermion in the ground state. Boson-boson 
and boson-fermion interaction energies have been denoted 
 by $E_{bb}$ and $E_{bf}$. 
The quantities $\Delta,\Delta'$ and $\Omega$ are given in terms of the 
boson/fermion densities $n_b(r), n_f(r)$ in the ground state by 
\begin{equation}
\label{DO}
  \Delta\equiv h\int\!\! d^3r \, r^2 
       \frac{dn_{f}(r)}{dr}\frac{dn_{b}(r)}{dr},\quad 
  \Delta'\equiv h\int\!\! d^3r \, r \left[ n_{f}(r)\frac{dn_{b}(r)}{dr}
       -\frac{dn_{f}(r)}{dr}n_{b}(r)\right] , \quad
  \Omega \equiv h\,\xi^2\int\!\! d^3r
          \frac{dn_{f}(r)}{dr}\frac{dn_{b}(r)}{dr}, 
\end{equation}
where $\xi=\sqrt{\hbar/m\omega_0}$. 
One may use the stationary condition of the ground state, 
\begin{equation}
    2\Ekin^+-2\Eho+3E_{bb}+3E_{bf}=0, \quad 
    2\Ekin^--2\Eho^-+3E_{bb}+\Delta'=0
\end{equation}
 in order to eliminate in eq.(\ref{egm}) the dependences on 
$E_{bb},E_{bf}$ and $\Delta'$. The monopole frequency is then rewritten as 
\begin{equation}
\label{egmm}
\frac{\omega_{M}(\theta)}{\omega_0}=\sqrt{5-\frac{\Ekin^{+}
+2\Ekin^{-}\cos{\theta}\sin{\theta}+2\Delta\sin^2{\theta}}
{\Eho^{+}+2\Eho^{-}\cos{\theta}\sin{\theta}}}.
\end{equation} 
We have checked that the Thomas-Fermi calculation of the ground 
state adopted below gives rise to a negligible difference if 
one evaluates either the  expression (\ref{egmm})  or (\ref{egm}). 


We calculate the ground state energies and densities of the 
boson-fermion mixed system in the Thomas-Fermi approximation which 
is valid for $gN_b\gg 1$ and $N_f\gg 1$ except around the 
surface region\cite{Molmer,Nygaard,Miyakawa,Amoruso}. 
We take harmonic oscillator length $\xi$ 
and energy $\hbar\omega_0$ as units, and define scaled dimensionless 
variables: the radial distance $x=r/\xi$, boson/fermion densities 
$\rho_{b,f}(x)=n_{b,f}(r)\xi^3/N_{b,f}$, and chemical potentials 
$\tilde{\mu}_{b,f}=2\mu_{b,f}/\hbar\omega_0$. We solve the coupled 
Thomas-Fermi equations, 
\begin{equation}
\label{tfeq}
      \tilde{g}N_{b}\rho_{b}(x)+x^2+\tilde{h}
      N_{f}\rho_{f}(x)=\tilde{\mu_{b}},  \quad
      [6\pi^2N_{f}\rho_{f}(x)]^{2/3}+x^2+\tilde{h}N_{b}\rho_{b}(x)
      =\tilde{\mu_{f}}, 
\end{equation}
where $\tilde{g}=2g/\hbar\omega_0\xi^3$ and 
$\tilde{h}=2h/\hbar\omega_0\xi^3$.

One of the most promising candidates for the realization of the 
mixed condensate is 
the potassium isotope system. Precise values of the scattering 
lengths are not well known at present and different values
have been reported \cite{Cote,Kscat1,Kscat2}.
We take for the boson-boson interaction  the parameters of the 
$^{41}$K-$^{41}$K system in \cite{Cote} and a trapping 
frequency of $450$Hz which gives $\tilde{g}=0.2$. For the 
boson-fermion interaction we take several values in the range 
$h/g=\tilde{h}/\tilde{g}=-2.37 \sim 3.2$. It should be noted that
the interaction strength can be controlled using 
Feshbach resonances\cite{Fesh}. 

We have performed a numerical calculation for $N_{b}=N_{f}=10^6$. 
In the ground state the fermions 
have a much broader distribution than bosons because of the Pauli 
principle. Fermions are further squeezed out of the center for a 
repulsive boson-fermion interaction ($h>0$) . They will eventually 
form a shell-like distribution around the surface of bosons 
for $h/g\ge 1$ and will be completely pushed 
away from the center ($n_{f}(0) = 0$) at around $h/g \sim 3$. 
For an attractive boson-fermion interaction, on the other hand, the 
central densities of the bosons and fermions increase together. 
The system becomes unstable against collapse at around 
$h/g= -2.37$ due to the strong attractive boson-fermion 
interaction\cite{Instbf}. 

 Figure 1 shows the kinetic energy, the oscillator potential energy, 
and the interaction energy contributions to the ground state energy 
 against the parameter $h/g$. The figure shows also 
the quantities $\Delta$ and $\Omega$ which represent the contributions 
of the boson-fermion interaction to the multipole frequencies 
(\ref{egm})-(\ref{egd}) and (\ref{egmm}). One may notice that the 
fermionic kinetic- and potential-energy contributions are 
a few times larger than the bosonic contribution 
in the present system. It is noted that $\Delta$ takes large negative 
values at both large negative and positive regions of $h/g$. In the 
former region the bose and fermi density distributions become coherent 
due to the attractive interaction, 
and the radial integral in eq.(\ref{DO}) takes a large positive value. 
In the opposite case ($h/g\gg 1$), the same integral changes sign 
because the fermions are pushed away from the center by the repulsive 
boson-fermion interaction, thus giving a large negative contribution 
in the bosonic surface region. In the region $0< h/g < 1$, on the 
other hand, the integral is slightly positive and $\Delta$ takes a 
small positive value. The quantity $\Omega$ follows the same trend as 
$\Delta$, but the absolute value 
is much smaller than $\Delta$, as the most important contribution to 
the integral comes from the surface region where $r\gg \xi$. 

Once the ground state parameters are 
obtained the frequencies (\ref{egm})-(\ref{egd}) are minimized against 
$\theta$.  
Usually, the sum rule approach predicts a strength-weighted average 
energy of eigenstates for a given multipolarity. The calculated energy 
 coincides with the true excitation energy if the 
relevant strength is concentrated in a single state. By adopting the 
minimization procedure we simultaneously determine the character of 
the low-lying collective mode  and the corresponding average energy. The 
character of the mode is given by the value of $\theta$, for instance, 
$\theta\simeq\pi/4$ for the bosonic- and $-\pi/4$ for the fermionic-modes, 
$\theta\simeq 0$ for the in-phase oscillation and $\pi/2$ for the 
out-of-phase oscillation of the two types of particles. 
As there are two kinds of particles involved in eq.(2), one would expect 
an emergence of two types of collective oscillations for each multipole. 
Another collective mode would have a character orthogonal to the low-lying 
mode as far as the phase relation of the two operators in (2) is concerned. 
In the present approach we calculate the frequency of the latter mode, 
the high-lying one, from eqs.(\ref{egm})-(\ref{egd}), by using the 
operator $F^+_\alpha \sin\theta_L-F^-_\alpha\cos\theta_L$ for each $\alpha$, 
where the mixing angle $\theta_L$ is the one determined for the low-lying 
mode. 

Figure 2 shows frequencies of the lower (solid lines) 
and the higher (dashed lines) modes for (a) monopole, (b) 
quadrupole  and (c) dipole cases as functions of $h/g$. 
The corresponding mixing angles $\theta$ determined by the minimization 
procedure for the lower mode are plotted in Fig.3 
for each multipolarity as a function of $h/g$. 
Below we discuss the behavior of the frequencies $\omega_\alpha$ 
by defining three regions of $h/g$: 
(I) $h/g<0$, (II) $0<h/g\lesssim 1$, (III) $1\lesssim h/g$.

\vspace{2mm}\noindent
{\it a)monopole}: \\
For a non-interacting boson-fermion system the low-lying monopole 
mode is the fermionic oscillation with  frequency 
$\omega_M^L=2\omega_0$, while the higher mode is the bosonic one 
with $\omega_M^H=\sqrt{5}\omega_0$ in the Thomas-Fermi approximation. 
Around  $h\simeq 0$ one may obtain 
$\omega^{L}_{M}\simeq 2 \omega_0 \left(1+A \tilde{h}N^{\frac{1}{6}}\right)$,
$\omega^{H}_{M}\simeq \sqrt{5} \omega_0 \left(1-A^{\prime}
\tilde{h}\tilde{g}^{-\frac{1}{5}}N^{\frac{3}{10}}\right)$, 
where $A=3\cdot 6^{\frac{1}{6}}/4 \sqrt{2}\pi^{2}$ and $A^{\prime} = 7 \sqrt{3}
/20\pi^{2} \cdot (8\pi/15)^{\frac{1}{5}}$. The mixing angle for the lower mode 
is given as $\theta_{M} = -\pi/4 - \delta_{M}$ 
with $\delta_{M} = (5\cdot 6^{\frac{1}{6}}/\sqrt{2}\pi^{2})
\tilde{h}N^{\frac{1}{6}}$. 
This behavior is seen in region (II) where the boson-fermion interaction 
is weakly repulsive and the lower monopole mode is of a fermionic 
character, see Fig.3 (solid line). 
Bosonic oscillation in this region is more rigid than the 
fermionic one because of the repulsive interaction among bosons. 
In region (I), the situation is quite different: The low-lying mode becomes 
a coherent boson-fermion oscillation as represented by the large negative 
value of $\Delta$, and the excitation energy shows a 
sharp decrease towards the instability point $h/g\sim -2.37$ of the 
ground state\cite{Instbf}, although $\omega_M$ does not become exactly zero 
within our approximation. 
In this region the attractive boson-fermion interaction is much more 
effective in the excited state than in 
the ground state and cancels out the increase in the kinetic energy. 
In region (III), too, we find that the low-lying mode becomes an in-phase 
oscillation. Here, the boson and the fermion densities in the ground 
state are somewhat separated, and the in-phase motion which keeps this 
separation is energetically more favorable than the out-of-phase motion 
as seen in the value of $\Delta$. 

\vspace{2mm}\noindent
{\it b)quadrupole}: \\
For the quadrupole excitation (Fig.2(b)), the lower 
(higher) energy mode is almost the pure  bosonic (fermionic) oscillation 
over the broad range of the $h/g$ values studied, Fig.2 (dashed line). 
To the first 
order in  $\tilde{h}$ the frequencies of the lower and the higher 
quadrupole modes are given by 
$\omega^{L}_{Q}\simeq \sqrt{2} \omega_0 \left(1-B 
\tilde{h}N^{\frac{1}{6}}\right)$,
$\omega^{H}_{Q}\simeq 2 \omega_0 (1-B^{\prime}
\tilde{h}\tilde{g}^{\frac{2}{5}}N^{\frac{7}{30}})$, 
where $B=6^{\frac{1}{6}}/4 \sqrt{2} \pi^{2}$ and $B^{\prime}=(15/8\pi)^{\frac{2}{5}}
/(14\cdot 6^{\frac{1}{6}}\sqrt{2}\pi^{2})$.
The corresponding mixing angle for the lower mode is given by 
$\theta_{Q} = \pi/4 + \delta_{Q}$ with 
$\delta_{Q} = 2^{3}B^{\prime} \tilde{h} \tilde{g}^{\frac{2}{5}}N^{\frac{7}{30}}$. 
For the quadrupole mode  similar mechanisms as for the monopole mode 
are at work concerning 
the dependence on $|h/g|$. The role of the boson-fermion 
interaction is however much reduced compared with the monopole case 
as seen by the factor 2/5 of eq.(\ref{egq}), which reflects that the 
quadrupole oscillation has five different components. Thus 
the quadrupole mode 
obtains an in-phase character only at large values of $|h/g|$. In the 
other region of $|h/g|$ the low-lying mode becomes a simple bosonic oscillation. 
This is because the fermionic mode costs a larger kinetic energy and 
favors $\theta=\pi/4$ as 
seen in the term $(\Ekin^b-\Ekin^f)\sin\theta\cos\theta$ in 
eq.(\ref{egq}).

\vspace{2mm}\noindent
{\it c)dipole}: \\
General arguments\cite{review} show that for a harmonic oscillator 
external potential a uniform shift of the ground state density 
generates an eigenstate of the system, corresponding to the boson-fermion 
in-phase dipole oscillation with frequency $\omega_0$. This is evident 
in Fig.2c) and also in eq.(\ref{egd}) at $\theta=0$. 
In the regions (I) and (III) the 
out-of-phase oscillation is unfavorable due to the same reason 
as for the monopole mode: It loses the energy gained in the 
ground state boson-fermion configuration.  
For a weakly repulsive 
$h$ an interesting possibility arises: In the region (II)
the out-of-phase mode of the boson-fermion oscillation lies 
lower than the in-phase mode. Let us first note that at $h=0$ 
the out-of-phase oscillation frequency becomes degenerate as the 
in-phase one because the bosonic and 
the fermionic dipole modes are independent. One may note that at 
$h/g\simeq 1$ again the degeneracy occurs. Here the potential term 
for the fermion becomes almost linear to the fermion density 
itself (see, eq.(\ref{tfeq})). This suggests that the fermion density is 
determined almost entirely by the chemical potentials and becomes 
nearly constant as far as the boson density is finite. A uniform dipole 
shift of fermions thus produces almost no effect on bosons, and results in 
the degeneracy of the frequency. Between 
$h/g=0$ and 1, the  boson-fermion repulsion is weaker for the out-of-phase 
oscillation than the in-phase one (and hence the ground state) as 
reflected in the sign of $\Omega$.

\vspace{2mm}
In the present paper, we studied  collective oscillations
of trapped boson-fermion mixed condensates using sum rule approach. 
We introduced a mixing angle of bosonic and fermionic multipole 
operators so as to study if the in- or out-of-phase motion of 
those particles is favored as a function of the boson-fermion interaction
strength.  For the monopole and quadrupole cases, the coupling of the 
bose- and fermi-type oscillation is not large for moderate values of 
the coupling strength $h$. At large values of $|h/g|$, the low-lying 
modes become an in-phase oscillation of bosons and fermions. This is 
especially so for the monopole oscillation at attractive boson-fermion 
interaction: The excitation energy of this mode almost 
vanishes at the instability point of the ground state. In the case 
of the dipole mode, in contrast, the in-phase oscillation 
remains an exact eigenmode with a fixed energy for harmonic 
oscillator potentials, while the average energy of the 
out-of-phase oscillation is strongly dependent on the 
boson-fermion interaction. We found that at weak repulsive values of 
the interaction the out-of-phase motion stays lower than the 
in-phase oscillation. 
In this paper we calculated also the frequencies of the high-lying modes 
for each multipole, by adopting the operators orthogonal to the low-lying 
modes. These modes, too, are collective in character and, in the present 
framework,  their average frequencies showed rather strong dependences on 
the boson-fermion coupling strength. 
Deeper insight into the collective modes 
studied in this paper will require a detailed investigation of 
the solutions of, e.g.,  the self-consistent RPA type equations 
that allow an arbitrary  radial dependence of the excitation 
operators.  Studies in this direction are now in progress. 

%
%

%
\newpage
%
\begin{figure}
\caption{(a) Ground state expectation values of the fermion kinetic 
energy $\Ekin^f$ (long dashed line), oscillator potential energies 
$\Eho^b$ (short-dash-dotted line) and $\Eho^f$ (long-dash-dotted line), 
boson-boson and boson-fermion interaction 
energies $E_{bb}$ (dashed line) and $E_{bf}$ (dotted line), and the 
quantity $\Delta$ (solid line) against the interaction 
strength ratio $h/g$. The values are given in the unit of 
$N\hbar\omega_0$ and are dimensionless.
(b) The quantity $\Omega$ in the same unit.}
\end{figure}
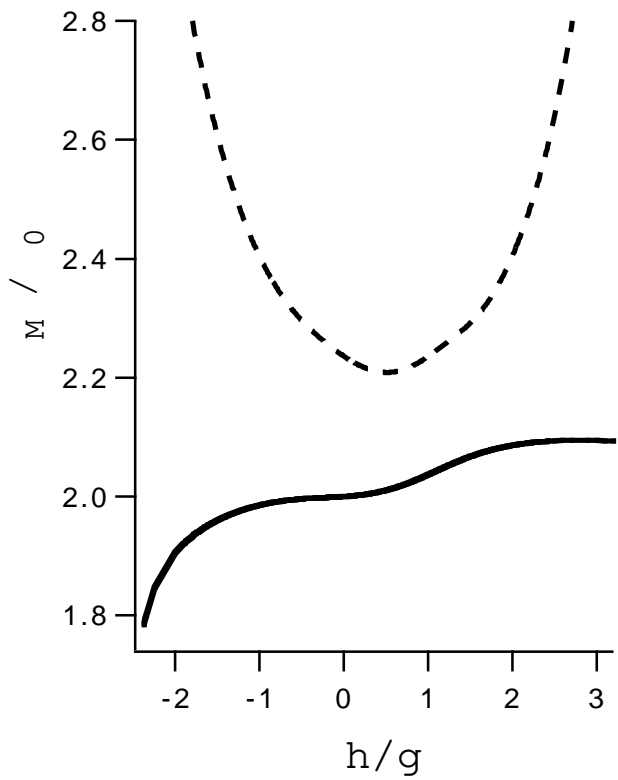
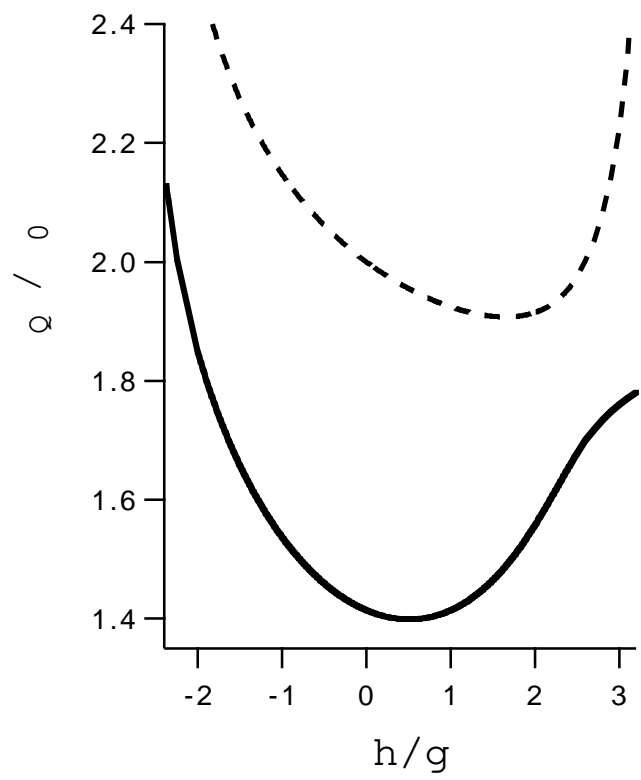
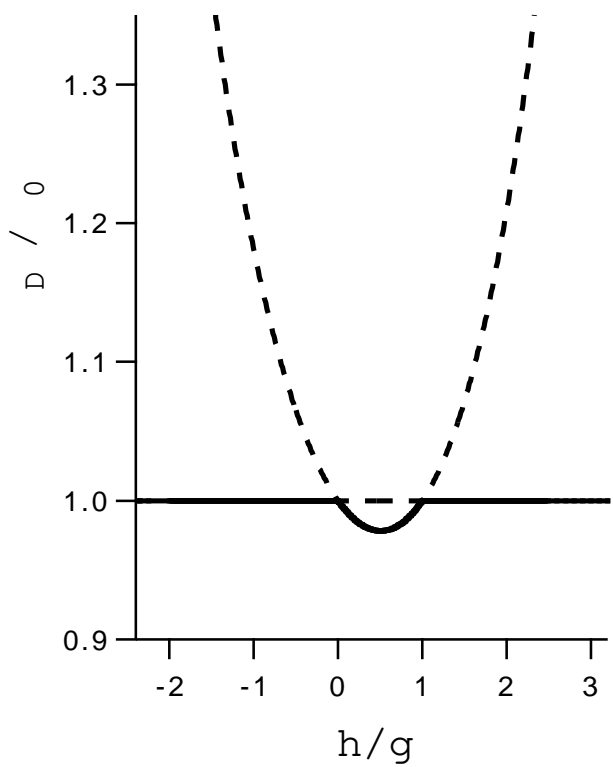
\begin{figure}
\caption{Excitation frequencies of collective oscillations
( a) monopole, b) quadrupole, c) dipole )
as  functions of $h/g=\tilde{h}/\tilde{g}$, 
calculated based on the eqs.(\ref{egm})-(\ref{egd}). The mixing angle 
$\theta$ has been determined so as to minimize $\omega$ for 
each multipole operator. The ordinates are given in the unit of 
$\omega_0$ and are  dimensionless. 
The solid (dashed) lines are the lower (higher) energy mode
for each oscillation.} 
\end{figure}

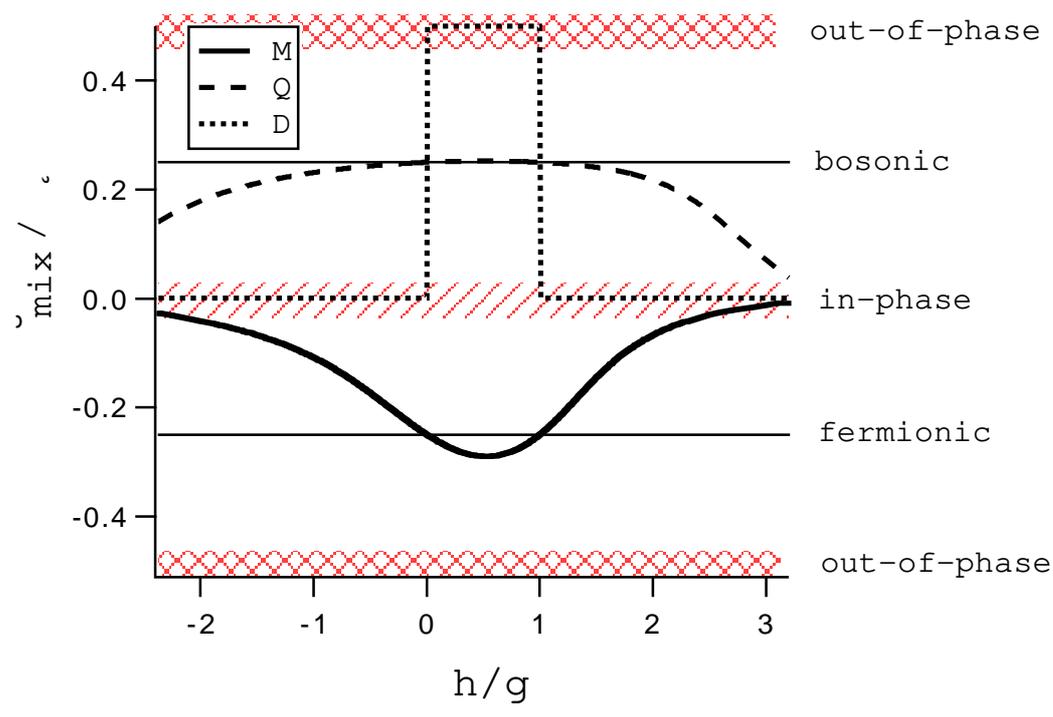
\begin{figure}
\caption{Mixing angles of in/out-of-phase excitation modes 
which are determined by minimizing excitation energies in Fig.3. 
The solid line shows the monopole, the dashed line
the quadrupole, and the dotted line the dipole oscillations. 
Shaded areas show  mainly in-phase (hatches) and 
mainly out-of-phase (cross-hatches) regions in $\theta$ 
Angles for  pure bosonic- or fermionic-modes are also indicated.}
\end{figure}

\begin{references}
\bibitem{review} 
For reviews, see: 
A. S. Parkins and H.~D.~F. Walls, Phys.~Reports {\bf 303}, 1 (1998);
F. Dalfovo, S. Giorgini, L.~P. Pitaevskii and S. Stringari, Rev.~Mod.~Phys.
{\bf 71}, 463 (1999).
%
\bibitem{Coll}
D.~S.~Jin, J.~R.~Encher, M.~R.~Mathews, C.~E.~Wieman and E.~A.~Cornell, 
Phys.~Rev.~Lett.{\bf 77},420(1996);
M.-O.~Mewes, M.~A.~Andrews, N.~J.~van~Druten, D.~M.~Kurn, D.~S.~Durfee, 
C.~G.~Tounsend and W.~Ketterle, Phys.~Rev.~Lett.{\bf 77},988(1996).
%
\bibitem{Stringari}
S.~Stringri, Phys.~Rev.~Lett. {\bf 77}, 2360 (1996).
%
\bibitem{Kimura}
T.~Kimura, H.~Saito and M.~Ueda, J.~Phys.~Soc.~Jpn.{\bf 68},1477(1999).
%
\bibitem{TBEC}
R.~Graham and D~Walls, Phys. Rev. {\bf 57A}, 484 (1998);
H~.Pu and N.~P.~Bigelow, Phys. Rev. Lett. {\bf 80}, 1134 (1998);
D.~Gordon and C.~M.~Sarvage, Phys. Rev. {\bf 58A}, 1440 (1998).
%
\bibitem{DeMarco}
B.~DeMarco and D.~S.~Jin,Science {\bf  285},  1703  (1999)
%
\bibitem{TFex}
L.~Vichi and S.~Stringari, Phys.~Rev.{\bf A60},4734(1999);
G.~M.~Bruun and C.~W.~Clark, Phys.~Rev.~Lett.{\bf 83},5415(1999);
M.~Amoruso, I.~Meccoli, A.Minguzzi and 
M. P. Tosi, Eur.~Phys.~J. {\bf D7},441(1999). 
%
\bibitem{symbf}
E.~Timmermans and R.~C\^{o}t\'{e}, Phys.~Rev. ~Lett. {\bf 80}, 3419 (1998);
W.~Geist, L.~You, and T.~A.~B.~Kennedy, Phys.~Rev.{\bf A59}, 1500 (1999);
M.-O.~Mewes, G.~Ferrari, F.~Schreck, A.~Sinatra and C.~Salomon, physics/9909007
%
\bibitem{hydro} 
I.~F.~Silvera, Physica {\bf 109 \& 110}, 1499 (1982); 
J.~Oliva, Phys.~Rev. {\bf B38}, 8811 (1988).
%
\bibitem{Molmer}
K.~M\o lmer, Phys. Rev. Lett.  {\bf  80},  1804  (1998)
%
\bibitem{Nygaard} 
N.~Nygaard and K. M{\o}lmer, Phys.~Rev. {\bf A59}, 2974 (1999).
%
\bibitem{Miyakawa}
T.~Miyakawa, K.~Oda, T.~Suzuki and H.~Yabu ; cond-mat/9907009.
%
\bibitem{Minniti} 
M.~Amoruso, C.~Minniti and M.~P.~Tosi, to be published in Eur.~Phys.~J. {\bf D}.
%
\bibitem{Instbf}
T.~Miyakawa, T.~Suzuki and H.~Yabu, cond-mat/0002048.
%
\bibitem{Amoruso} 
M.~Amoruso, A.~Minguzzi, S.~Stringari, M.~P.~Tosi and L.~Vichi,
Eur.~Phys.~J. {\bf D4}, 261 (1998);  
L.~Vichi, M.~Inguscio, S.~Stringari and G.~M.~Tino, 
J.~Phys. {\bf B31}, L899 (1998). 
%
\bibitem{tsurumi}
T.~Tsurumi and M.~Wadati, J.Phys.~Soc.~Jpn.{\b 69} (2000) in print.
%
\bibitem{Cote}
R.~Cot\'e, A.~Dalgarno, H.~Wang and W.~C.~Stwalley, 
Phys.~Rev. {\bf A57}, R4118 (1998). 
%
\bibitem{Kscat1} 
J.~L.~Bohn, J.~P. Burke,Jr, C.~H. Greene, H.~Wang, P.~L. Gould and 
W.~C. Stwalley, Phys. Rev. {\bf A59}, 3660  (1999). 
%
\bibitem{Kscat2}
B. Demarco, J.~L. Bohn, J.~P. Burke, Jr,
M. Holland and D.~S. Jin, Phys. Rev. Lett. {\bf 82}, 4208  (1999).
%
\bibitem{Fesh} 
E. Tiesinga, A.~J. Moerdijk, B.~J. Verhaar and H.~T.~C. Stoof,
Phys.~Rev.{\bf A46}, R1167 (1992); 
E. Tiesinga, B.~J. Verhaar, and H.~T.~C. Stoof, 
Phys.~Rev.{\bf A47}, 4114 (1993); 
J.~L. Bohn, cond-mat/9911132.
%
\end{references}
\end{document}